\documentclass[12pt]{article}
\usepackage{a4wide}
\usepackage{graphicx}

\newcommand{\slk}{/\kern-6pt k}
\newcommand{\sll}{/\kern-4pt l}
\newcommand{\slp}{p\kern-5pt/}
\newcommand{\slq}{q\kern-5.5pt/}
\newcommand{\sls}{s\kern-5.5pt/}
\newcommand{\slv}{v\kern-5pt\raise1pt\hbox{$\scriptstyle/$}\kern1pt}

\newcommand{\MeV}{{\rm\,MeV}}

\newcommand{\bbbone}{\hbox{\rm 1\kern-3pt l}}

\begin{document}

\begin{center}
{\Large\bf Decay and structure of heavy flavour}\\[1.3cm]
{\Large Stefan Groote, Arpan Chatterjee and Maria Naeem\footnote{Talk given
by Maria Naeem at SHARP 2026, 4.-7.\ March 2026, Curia, Portugal}}\\[12pt]
{Institute of Physics, University of Tartu,
W.~Ostwaldi~1, EE-50411 Tartu, Estonia}
\end{center}
\vspace{1cm}
\begin{abstract}\noindent
In accordance to the aim of the constituing meeting of the COST action CA24159
``Structure and Spectroscopy of Hadrons Research Project'' to introduce the
different groups to the action, in this talk we give an overview over the
subjects dealt with by the working group in Tartu related to hadron physics.
We deal with the production and the nonleptonic decays of charmed baryons in
the framework of the current algebra approach in terms of tensor invariants
and explain how this approach can be used to approach CP violation via long
distance effects in rescattering. New physics effects can be even seen in the
classical neutron beta decay, but the helicity approach used here is also
useful for e.g.\ calculating first order electroweak radiative corrections to
the decay of the polarised $W$ boson. Identical particle and mass effects are
seen in the Higgs decay into four leptons of the same type. The second main
part starts with indications for the intrinsic charm mechanism, explaining
the discrepancy between the results of SELEX and LHCb. The solution offered
here is valid only if one considers nonlocal field operators. The nonlocal
extension of the Nambu--Jona-Lasinio model, derived directly from QCD and
combined with the relativistic Faddeev approach, allows for the description
of hadronic states. We conclude by presenting open questions to the action.
\end{abstract}

\newpage

\section{Introduction}
The small working group in Tartu as part of the Laboratory of Theoretical
Physics dealing with theoretical particle physics consists in moment of
Assoc.\ Prof.\ Dr.\ Stefan Groote and three PhD students: Maria Naeem from
Pakistan, Arpan Chatterjee from India, and Ernest Michael Priidik Gallagher
from Estonia, the latter only partially involved in particle physics. The
working group was associate member of the COMPASS experiment (COmmon Muon and
Proton Apparatus for Structure and Spectroscopy, MoU signed in 2018) and is
full member of the AMBER experiment (Apparatus for Meson and Baryon
Experimental Research, MoU signed in 2024). The group leader is active member
in the CERN Baltic Group (Study Group) and is teaching theoretical particle
physics in Estonian and English for bachelor and master students at the
University of Tartu. In the past, the group has dealt with research on Heavy
Flavour Physics in many different collaborations.

The talk is organised as follows: In Section~2 we present our ideas on the
production and nonleptonic decays of charmed baryons. Section~3 contains the
main subject of the presenter. In Section~4 we introduce the intrinsic charm
mechanism dealt with mainly with the former PhD student Sergey Koshkarev,
and Section~5 contains the nonlocal extension of the Nambu--Jona-Lasinio
model, the subject of Arpan Chatterjee. We conclude with presenting open
questions.

\section{Production and decays of charmed baryons}
In the advent of heavy baryon data from LHCb, Belle~II and BESIII (for a
recent overview, see Ref.~\cite{Li:2025nzx}), we concentrate on the production
and nonleptonic decays of charmed baryons. For the production of charmed
$\Lambda$ baryons at $e^+e^-$ colliders, we put particular emphasis on mass
and polarisation effects, as it is expected that the masses and polarisation
states of the heavy partons produced at these colliders will be transferred to
the final channel. Heavy baryons like these can decay with and without leptons
in the final channel. We apply the current algebra approach to nonleptonic
decays of charmed baryons.
\begin{figure}[t]\begin{center}
\includegraphics[scale=0.2]{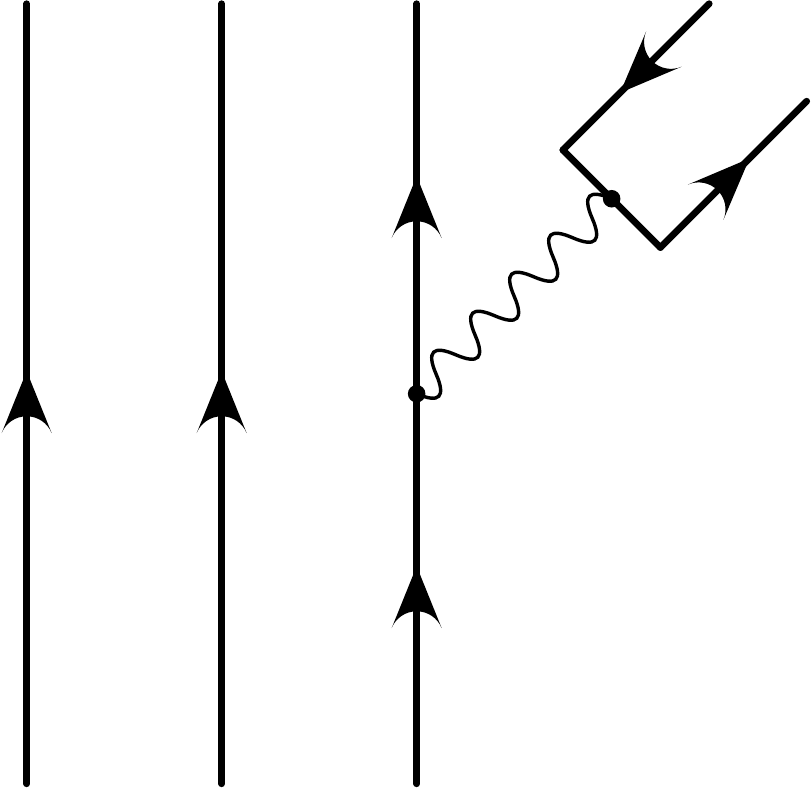}\
\includegraphics[scale=0.2]{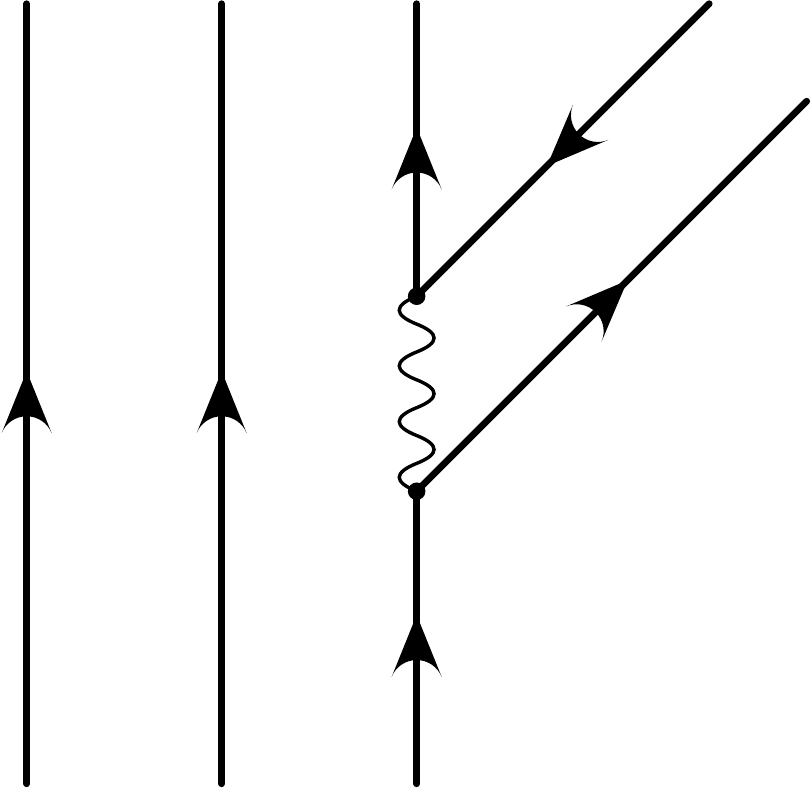}\
\includegraphics[scale=0.2]{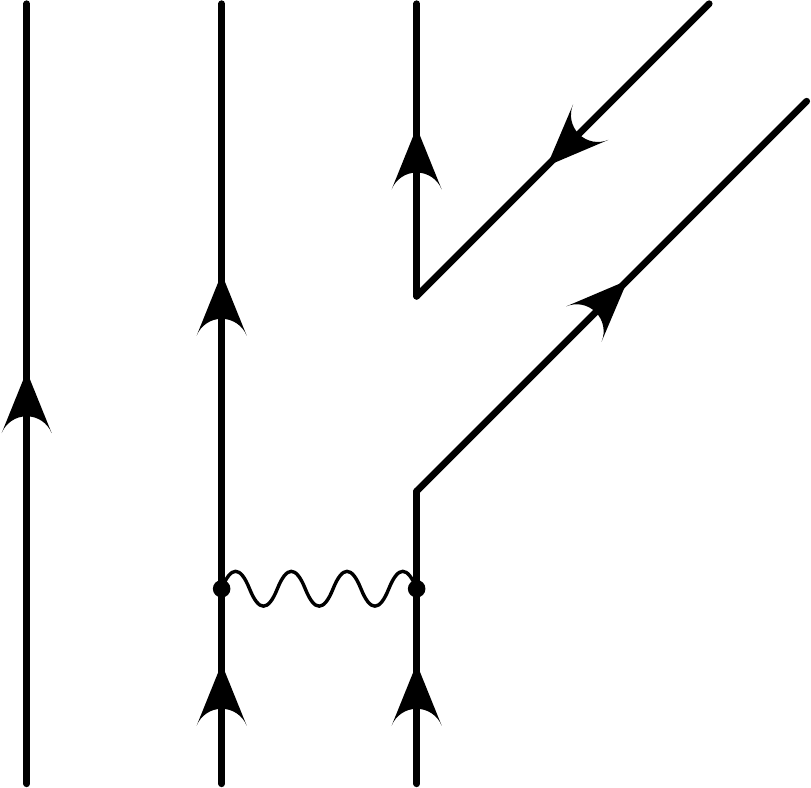}\
\includegraphics[scale=0.2]{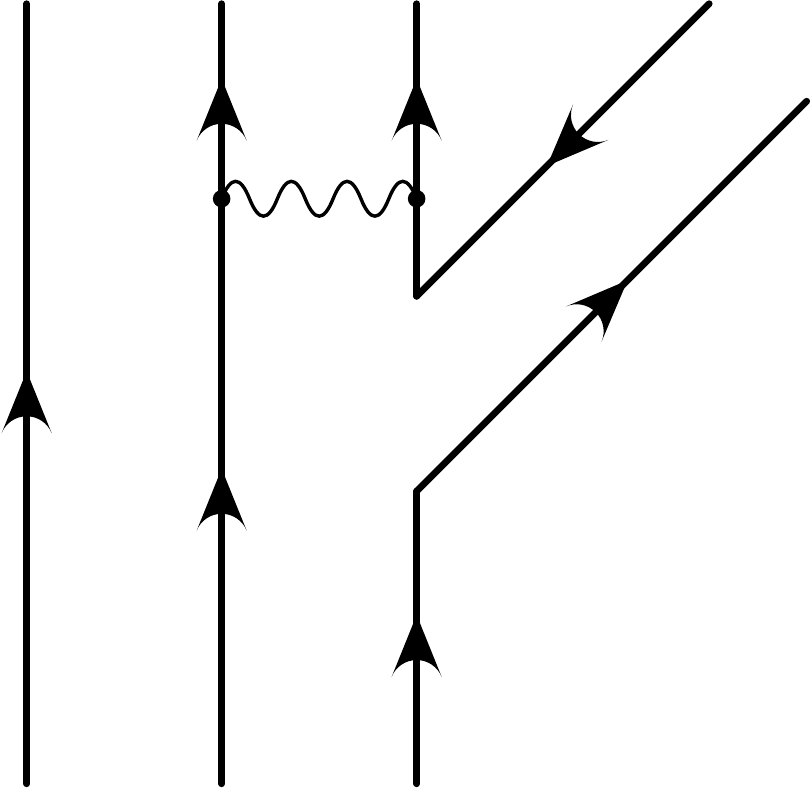}\
\includegraphics[scale=0.2]{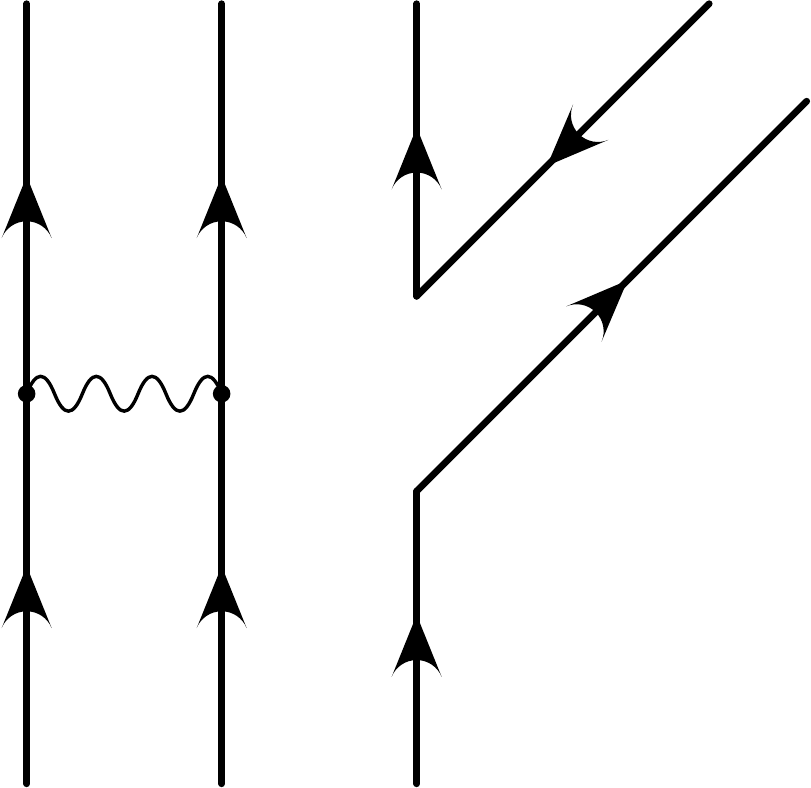}
\caption{\label{topo}Diagrams contributing to nonleptonic decays of
charmed baryons}
\end{center}\end{figure}
In Figure~\ref{topo} the diagrams are shown that contribute to the Hamiltonian
$H$ of the decay processes. The transition amplitudes
\begin{equation}
  {\cal M}(B_i \stackrel{H}{\longrightarrow}B_fM_k)={\cal M}_{fki}
  =\sum_jI^j_{fki}{\cal T}_j
\end{equation}
of the decay process are expressed in termd of 7 tensor invariants
${\cal T}_j$, as described in Ref.~\cite{Groote:2021pxt}. As further plans in
this context, together with colleagues from Uppsala and Warsaw (Andrzej Kupsc
(SE), Patrik Adlarson (SE) and Artur Ukleja (PL)) we plan to extend the
calculations in the following directions:
\begin{itemize}
\item CP violation in charmed baryon decays via rescattering,\\
instead of short distance effects providing long distance
effects~\cite{He:2024unv}, and
\item extension of the two-body decay to a three-body decay\\
through a subsequent strong decay of the baryon--meson pair.
\end{itemize}
With this, we aim
\begin{itemize}
\item to explain a CP violation capable to trigger the baryon asymmetry, and
\item to model decay channels for e.g.\ $\Xi_c\to pK\pi$ seen at LHCb.
\end{itemize}
As a further branch of our research related to decays of baryons, together
with Bla\v zenka Meli\'c from Zagreb we plan to work on new physics effects in
the neutron beta decay by using the helicity approach~\cite{Groote:2019rmj}.

\section{Electroweak radiative corrections}
\begin{figure}[ht]\begin{center}\noindent
\includegraphics[scale=0.2]{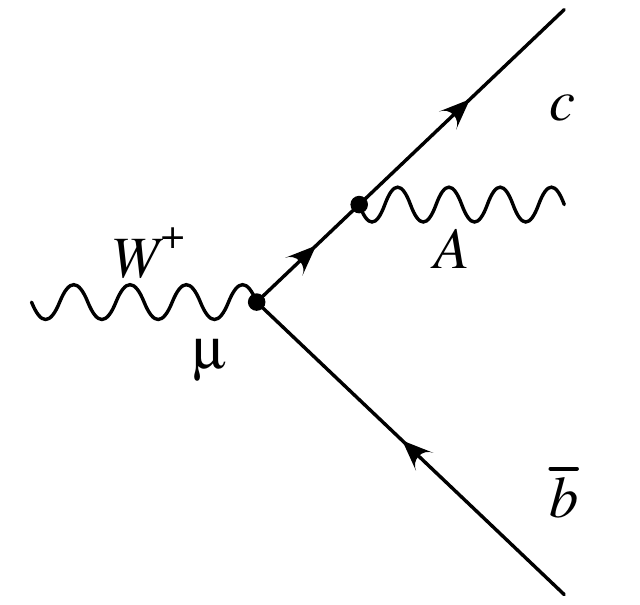}\
\includegraphics[scale=0.2]{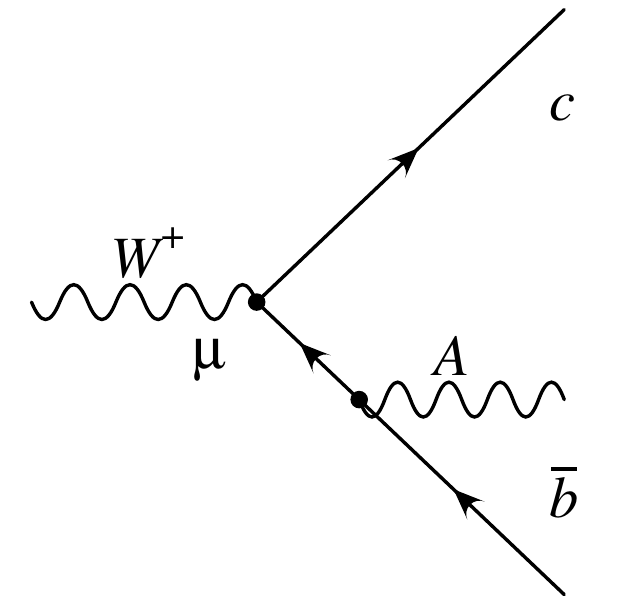}\
\includegraphics[scale=0.2]{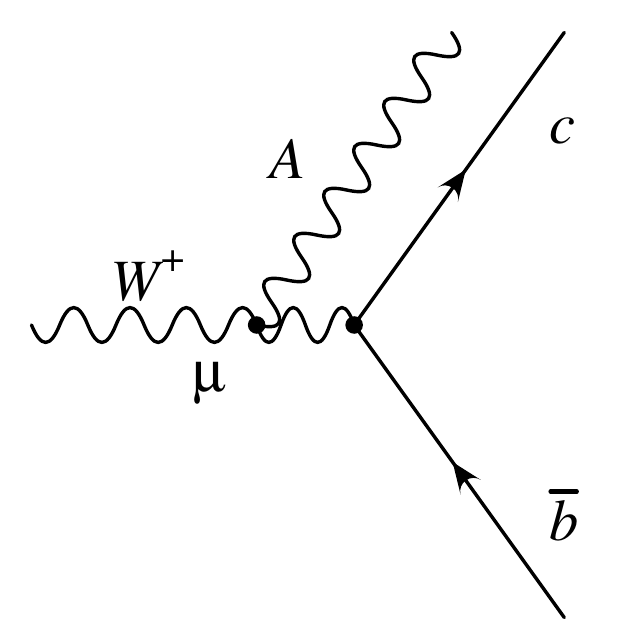}\quad
\includegraphics[scale=0.2]{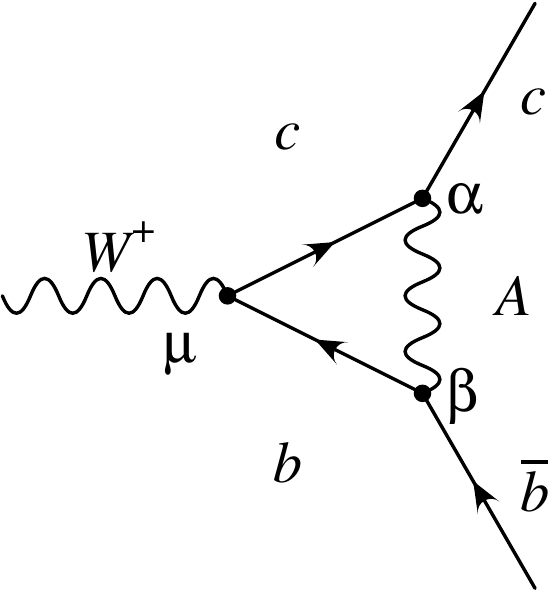}\quad
\includegraphics[scale=0.2]{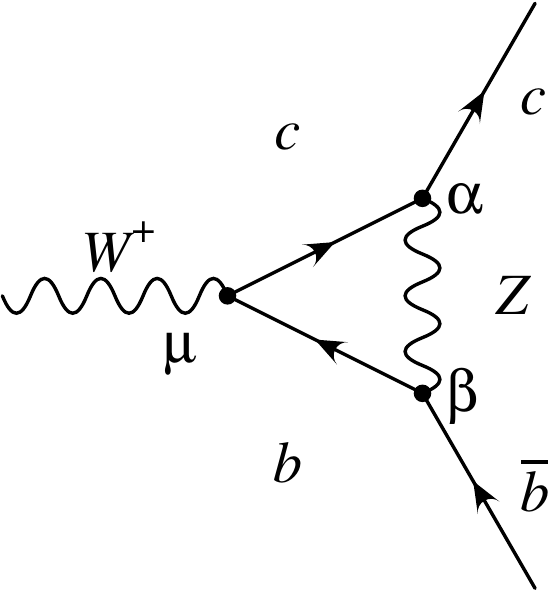}\quad
\includegraphics[scale=0.2]{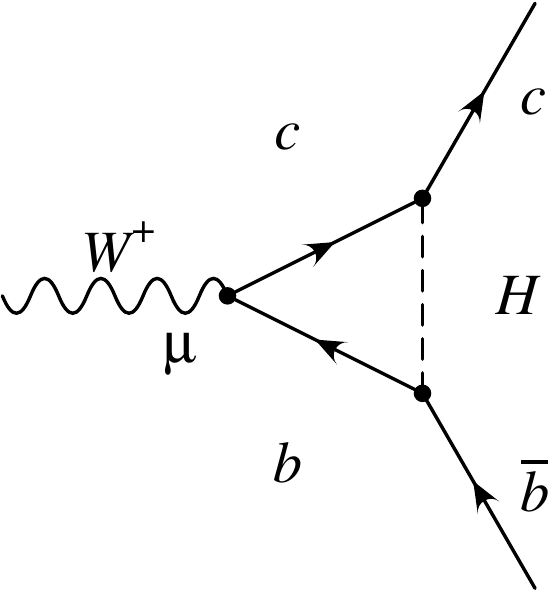}\quad
\includegraphics[scale=0.2]{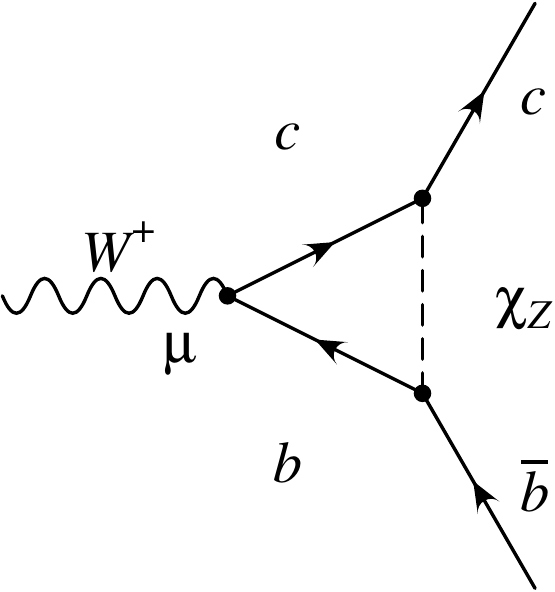}\\[7pt]\strut\quad
\includegraphics[scale=0.2]{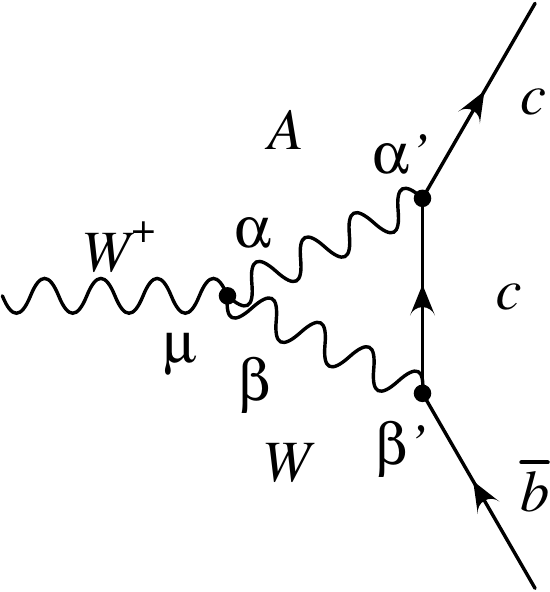}\quad
\includegraphics[scale=0.2]{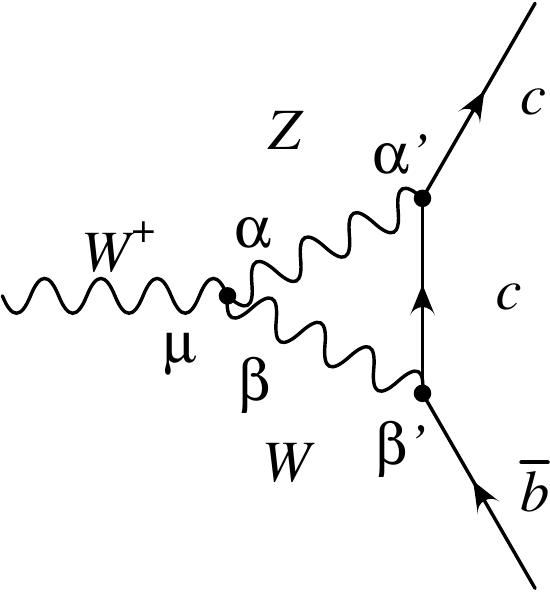}\quad
\includegraphics[scale=0.2]{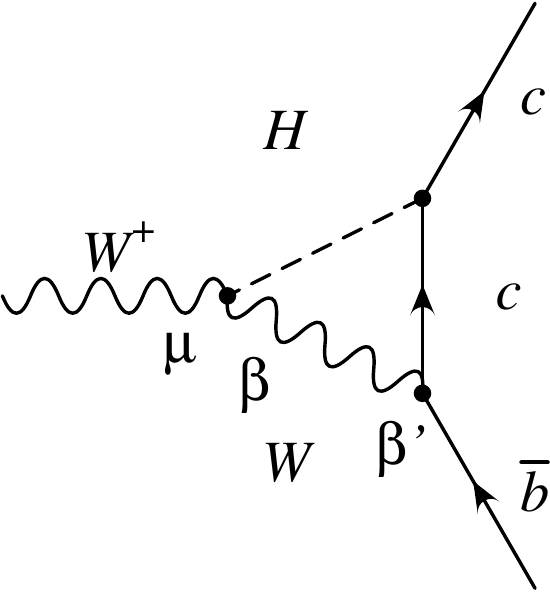}\quad
\includegraphics[scale=0.2]{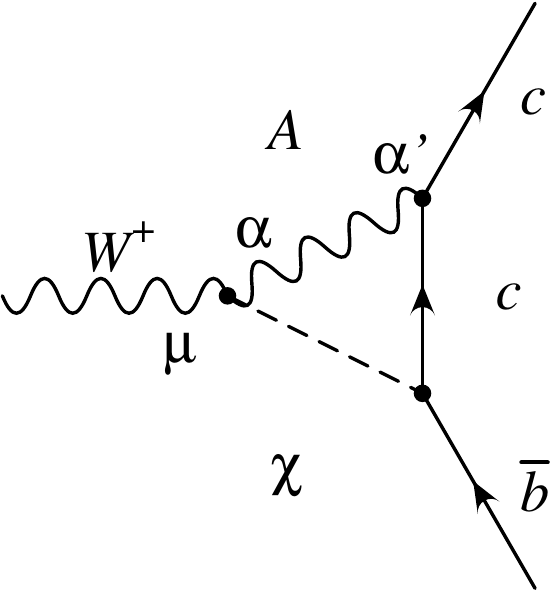}\quad
\includegraphics[scale=0.2]{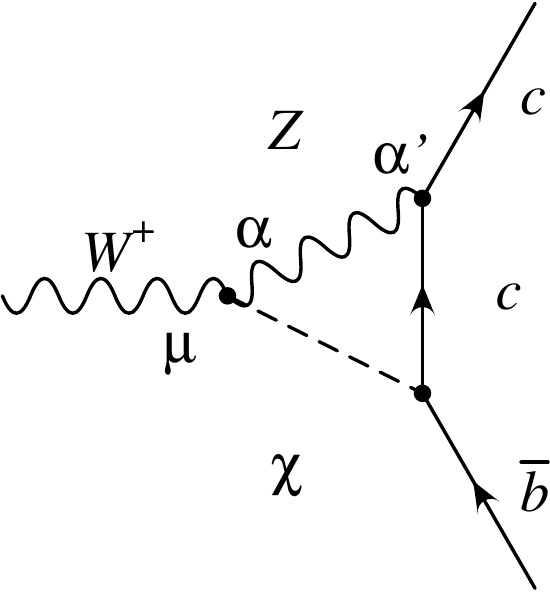}\quad
\includegraphics[scale=0.2]{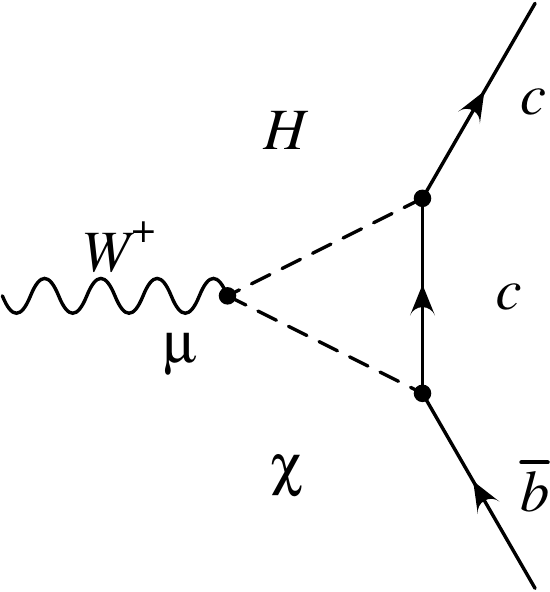}\quad
\includegraphics[scale=0.2]{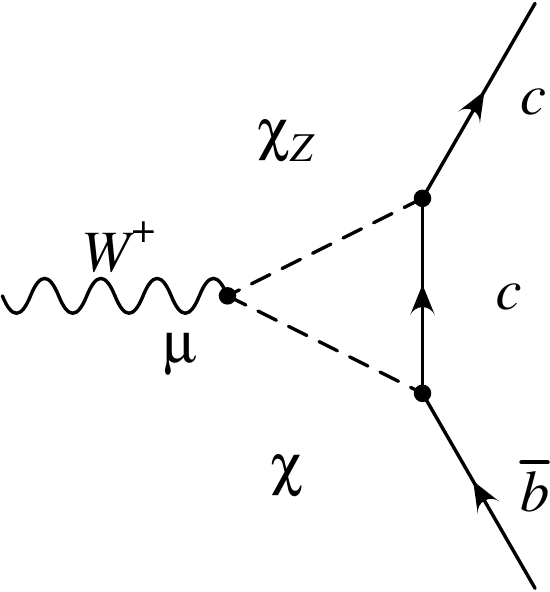}\\[7pt]\strut\quad
\includegraphics[scale=0.2]{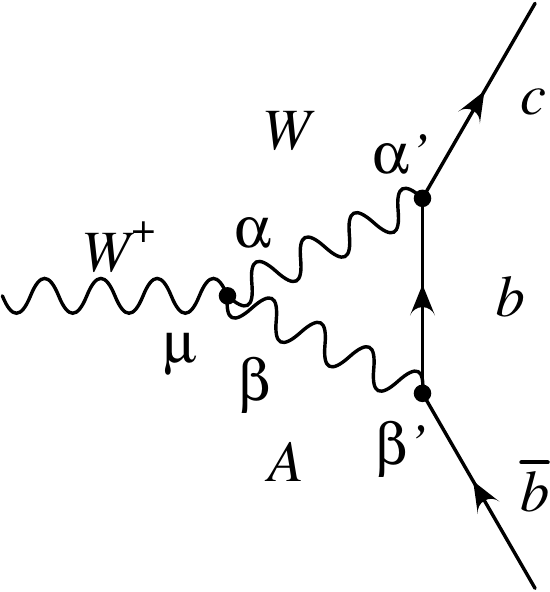}\quad
\includegraphics[scale=0.2]{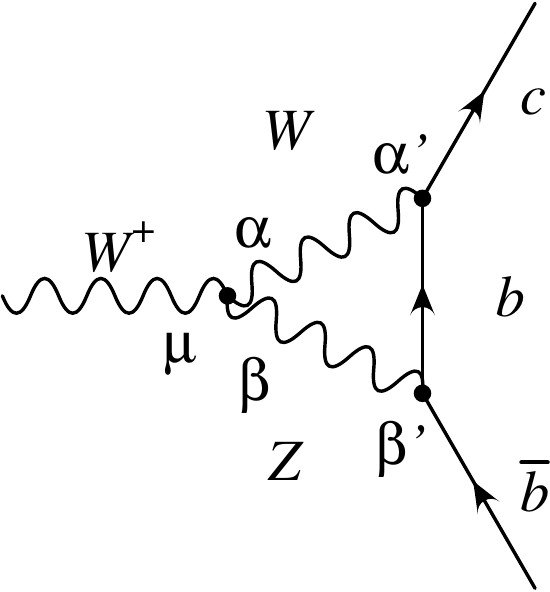}\quad
\includegraphics[scale=0.2]{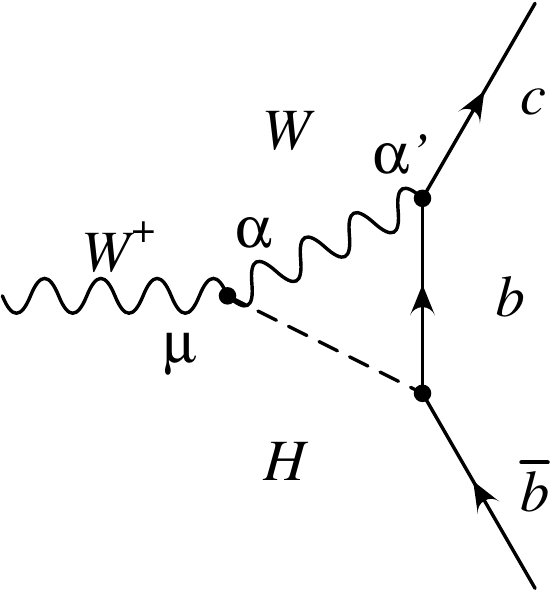}\quad
\includegraphics[scale=0.2]{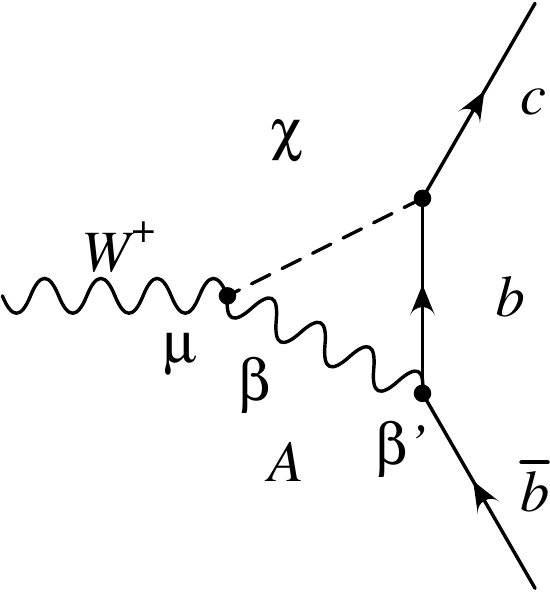}\quad
\includegraphics[scale=0.2]{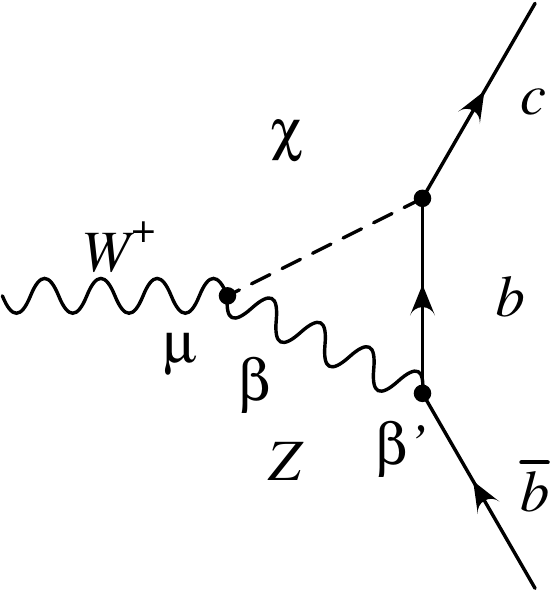}\quad
\includegraphics[scale=0.2]{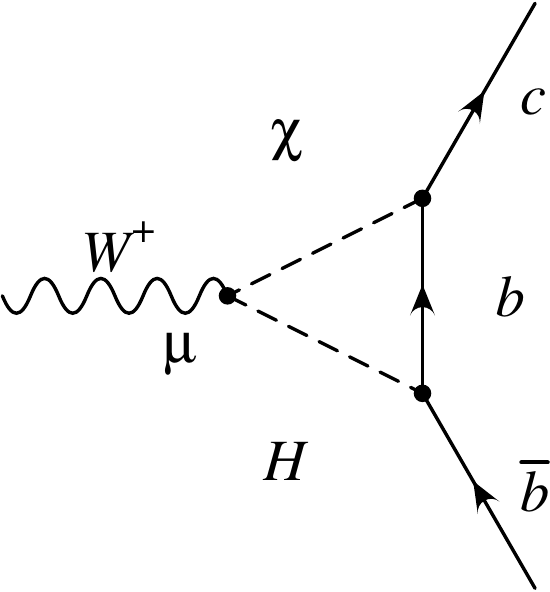}\quad
\includegraphics[scale=0.2]{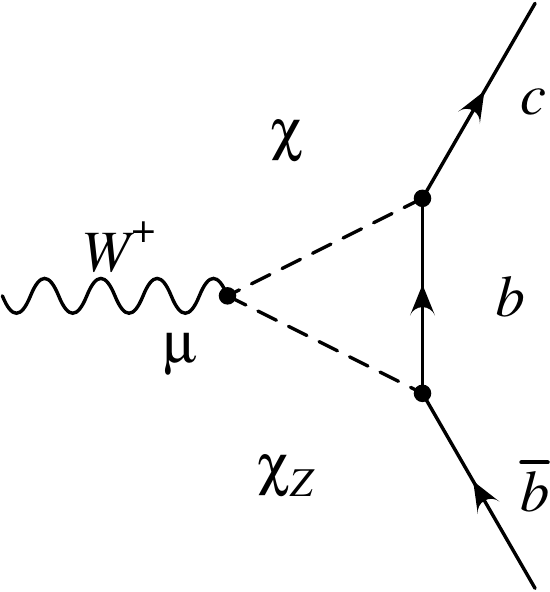}
\caption{\label{wewpol}Diagrams contributing to NLO electroweak corrections}
\end{center}\end{figure}

\begin{figure}[ht]\begin{center}
\includegraphics[scale=0.4]{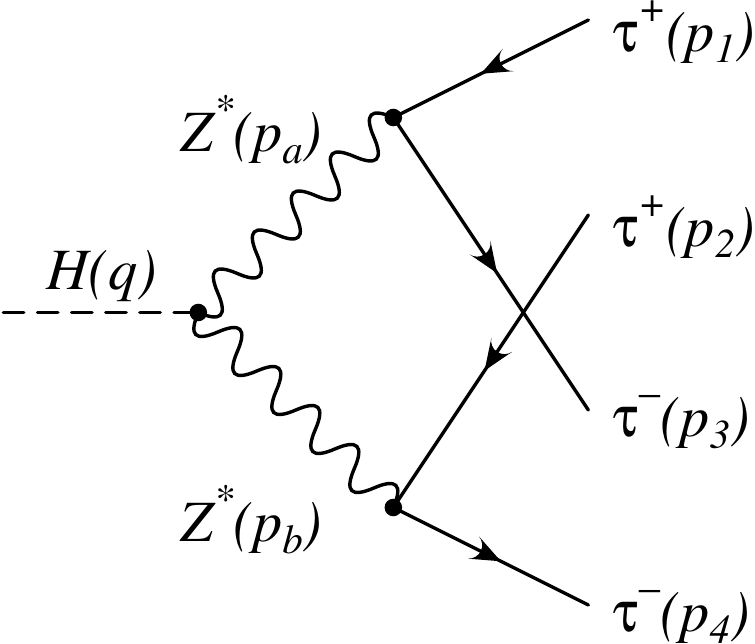}\quad
\includegraphics[scale=0.4]{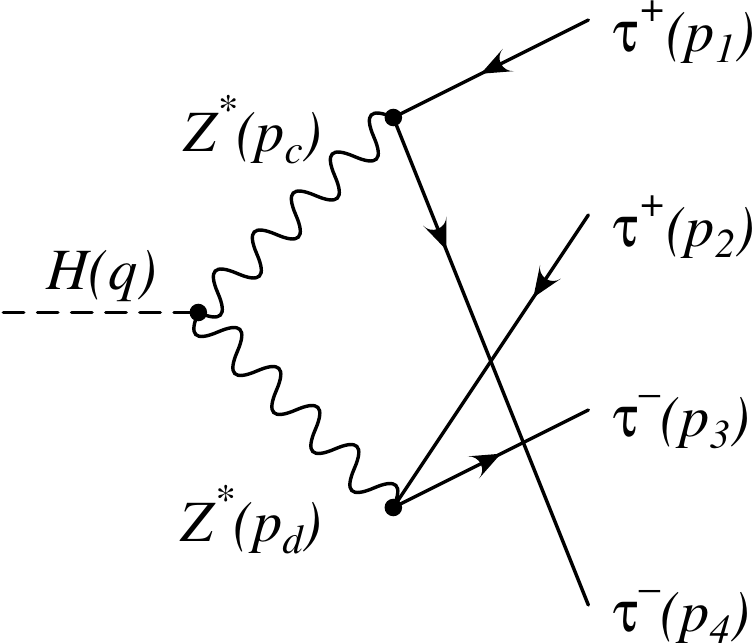}
\caption{\label{higgstau}Mixing contributions for
$H\to Z^\ast(\to\ell^+\ell^-)+Z^\ast(\to\ell^+\ell^-)$}
\end{center}\end{figure}

The top quark is by far the heaviest fermion found at the LHC. As the top
quark does not hadronise, ths fermion provides a ``clear'' state. The main
decay channel $t\to W+b$ that gives a polarisation to both of the final
states is analysed in detail in the past. The cascade is continued by the
decay of the $W$ boson. While QCD corrections to hadronic decays like
$W\to c+\bar b$ have been dealt with already, the main subject of the PhD
thesis of the presenter of this talk is the calculation of first order
electroweak radiative corrections to the decay of the polarised $W$ boson.
The diagrams contributing to this process are shown in Figure~\ref{wewpol},
and the results are published in Ref.~\cite{Naeem:2024wfw}.
In addition to mass effects considered in this reference, identical particle
effects in the ``golden Higgs channel''
$H\to Z^\ast(\to\ell^+\ell^-)+Z^\ast(\to\ell^+\ell^-)$ are considered in
Ref.~\cite{Groote:2022pjv}. Mass effects for the heaviest lepton $\tau$ amount
to $10\%$ and are non-negligible, while identical particle effects lead to a
mixing of the contributions shown in Figure~\ref{higgstau}.
The quantity that finally can be measured is the mean differential decay rate
\begin{equation}
  \frac{d\Gamma}{dp^2d\cos\theta}=\frac14\left(
  \frac{d\Gamma}{dp_a^2d\cos\theta^a}+\frac{d\Gamma}{dp_b^2d\cos\theta^b}
  +\frac{d\Gamma}{dp_c^2d\cos\theta^c}+\frac{d\Gamma}{dp_d^2d\cos\theta^d}
  \right).
\end{equation}

\section{Indications for the intrinsic charm mechanism}
\begin{figure}[ht]\begin{center}
\includegraphics[scale=0.25]{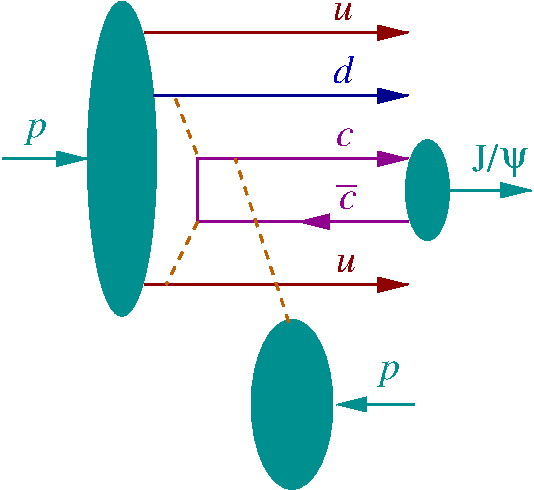}
\caption{\label{pppsi}Production of $J/\psi$ in proton--proton scattering via
the IC mechanism}
\end{center}\end{figure}

The intrinsic charm (IC) mechanism is a rigorous prediction of Quantum
Chromodynamics (QCD)~\cite{Brodsky:1980pb}. It is base on a prepared Fock
state for the proton,
\begin{equation}
|p\rangle\sim|uud\rangle+|uudG\rangle+|uudc\bar c\rangle+\ldots,
\end{equation}
where the fluctuation occuring at higher twist in the operator product
expansion is proportional to the inverse of the squared heavy mass $1/m_Q^2$.
The miscroscopically generated Fock state is minimal offshell and allows for
a soft scattering boosted close to the beam axis, as shown for the case of
the production of a single $J/\psi$ in Figure~\ref{pppsi}.
The final $c\bar c$ state carries a large portion of momentum, expressed by a
high Feynman $x_F$, which is characteristic for production of heavy flavour
via the IC mechanism, and is observable only in fixed target experiments like
NA3, SELEX, COMPASS or AMBER. For the future we plan a simulation to be
included in FLUKA at CERN.

The IC mechanism also nicely explains a discrepancy unsolved so far between
the mass states for an isospin pair of doubly charmed baryons. While in 2002
SELEX has measured the mass of the $\Xi_{cc}^+$ to be $3520\MeV/c^2$, LHCb in
2017 has found that the mass of the isospin partner $\Xi_{cc}^{++}$ is
$3621\MeV/c^2$. The mass difference is not appropriate for an isospin pair.
The solution offered in Ref.~\cite{Brodsky:2017ntu} is expressed in terms of
spin-0 and spin-1 diquark states $[\ ]$ and $(\ )$, respectively. While
$|\Xi_{cc}^+\rangle=|[dc]c\rangle$ could be produced via the IC mechanism and
is visible only at fixed target experiments, due to the higher spin state,
$|\Xi_{cc}^{++}\rangle=|u(cc)\rangle$ produced via gluon--gluon fusion can
have a much higher mass.

\section{Nonlocal extension of the NJL model}
The Fierz rearrangement that would allow for transitions between the afore
mentioned states holds only in case of local field operators. Based on works
of Diakonov and Petrov~\cite{Diakonov:1983hh,Diakonov:1987ty} and examplified
in the instanton liquid model~\cite{Anikin:2000rq}, a nonlocal extension of
the Nambu--Jona-Lasinio (NJL) model is
proposed~\cite{Plant:1997jr,Bowler:1994ir} which, besides being renormalisable
and providing confinement for the quarks, results in a compact
baryon~\cite{Rezaeian:2004nf}. In the PhD thesis of Arpan Chatterjee and in
close collaboration with Marco Frasca and Anish Ghoshal, we are able to derive
the nonlocal NJL model directly from QCD~\cite{Frasca:2021mhi}. The starting
point is the QCD lagrangian
\begin{equation}
  {\cal L}_{\rm QCD}=\bar\psi(i\gamma^\mu\partial_\mu-m_q
  +g\gamma^\mu A_\mu^aT_a)\psi-\frac14F_{\mu\nu}^aF^{\mu\nu}_a,
\end{equation}
leading to the Euler--Lagrange equations
$(i\gamma^\mu\partial_\mu-m_q+g\gamma^\mu A_\mu^aT_a)\psi(x)=0$ and
$(\partial_\mu\partial^\mu-M_g^2)A^\nu_a(x)=g\bar\psi(x)\gamma^\nu T_a\psi(x)$.
The latter is solved by
\begin{equation}
A^\nu_a(x)=\int G(x-y)g\bar\psi(y)\gamma^\nu T_a\psi(y)d^4y,
\end{equation}
where $(\partial_\mu\partial^\mu-M_g^2)G(x-y)=\delta^{(4)}(x-y)$. Inserting
one into the other, one ends up with a nonlocal NJL lagrangian
\begin{eqnarray}
  {\cal L}_{\rm NJL}&=&\bar\psi(x)(i\gamma^\mu\partial_\mu-m_q)\psi(x)
  \nonumber\\&&\strut+g^2\bar\psi(x)\gamma^\mu T_a\psi(x)\int G(x-y)
  \bar\psi(y)\gamma_\mu T_a\psi(y)d^4y.\qquad
\end{eqnarray}
In order to describe baryonic states, in the relativistic Faddeev approach the
three-body problem of the baryon is reduced to an effective two-body problem,
describing the baryon as a bound state of a diquark and a spectator quark. The
diquark is not a free, asymptotic state but an integral part of the baryon.
Along the lines given by the local NJL model~\cite{Oettel:2001kd}, the
calculation is performed in the following steps:
\begin{enumerate}
\item solution of the Bethe--Salpeter equation for the diquark,
\item scalar and axial diquarks expressed in terms of covariants,
\item wave functions expanded in terms of Chebychev polynomials,
\item numerical solution of the Faddeev equation for the baryon.
\end{enumerate}

\section{Conclusions}
As future goals, we offer to put more emphasis on non-perturbative methods
to explain
\begin{itemize}
\item CP violation for baryonic states
\item lowering of ground state masses
\item explanation of baryon confinement
\end{itemize}
Our group is one of the investigators in a Teaming For Excellence application
this year. Participants are TalTech and the National Institute of Chemical
Physics and Biophysics (NICPB) in Tallinn, the University of Tartu, CERN, and
the Helsinki Institute of Physics (HIP). The planned budget for five years is
12 Million Euro, and the sustainable aim is to build up a Center for Long-term
Excellence in Estonia (CIRCLE) at the example of the HIP in Finland.

\end{document}